# A First-principles Study of Weyl Nodal Loop and Multiple Sets of Weyl Points in Trigonal $PtBi_2$

Lin-Lin Wang[1,2*]

[1]Ames National Laboratory, Ames, IA 50011, USA

[2]Department of Physics and Astronomy, Iowa State University, Ames, IA 50011, USA

*llw@ameslab.gov

# Abstract

Coexistence of surface superconductivity and Fermi arcs in trigonal $\gamma$-$PtBi_2$ has recently attracted attention for possible realization of topological superconductivity. The Fermi arcs on the two different (0001) surface terminations have been associated with the set of Weyl points just above the Fermi energy ($E_F$). Here using first-principles calculations to explore the band crossings over the full Brillouin zone between the nominally highest valence and lowest conduction bands in $\gamma$-$PtBi_2$, we study the Weyl nodal loop (WNL) and multiple sets of Weyl points (WPs). The main difference between the two reported experimental structural parameters is the magnitude of Bi-layer buckling. While the WNL, bulk gap region and the set of Weyl points just above the $E_F$ are robust, the number and location of the other sets of WPs depend sensitively on the structural parameters with different magnitude of Bi-layer buckling. Besides calculating the 2D Fermi surface with Fermi arcs and quasi-particle interference (QPI) around the $E_F$ in good agreements with ARPES and experimental QPI, we also predict new Fermi arc features at higher energy.

## I. Introduction

Trigonal $\gamma$-$PtBi_2$[1] has shown large linear magnetoresistance[2-4] and its electronic band structure has been studied in earlier works for triply degenerated point[3], large Rashba-like spin splitting[5] and other interesting surface states[6-8]. More recently the coexistence of surface superconductivity (SC) and Fermi arcs (FAs) on trigonal $\gamma$-$PtBi_2$ (0001) surface has attracted much attention for possible realization of topological superconductivity[9-25]. Surface SC of $\gamma$-$PtBi_2$ has manifested in the SC gap opening[9, 11, 20-22] in scanning tunneling spectroscopy (STS) and contact SC[14, 23] with STS tips, as well as the sharpening of FAs[10, 18] in angle-resolved photo-emission spectroscopy (ARPES). A sizable SC gap opening of a few meV has been reported in a range from 2 K[11, 20, 22] to as high as 45 K[21]. A very recent low-temperature STS study[20] has discovered the SC vortex under magnetic field with high mobility to prove the 2D surface SC in $\gamma$-$PtBi_2$. The pairing mechanism and possible Majorana particles are still being actively investigated[19, 24, 26, 27]. The FAs on the two different (0001) surface terminations have been associated with the set of inversion-breaking Weyl points (WPs) just above the Fermi energy ($E_F$) [10, 16, 28]. A recent study has found that the Weyl nodal loop (WNL) in $\gamma$-$PtBi_2$ has transport signatures of anomalous planar Hall effect and a small magnetic field can induce additional WPs[29]. Another theoretical study proposed that a monolayer of $\gamma$-$PtBi_2$ can behave as a 2D topological ferroelectric metal[30].

Here using density functional theory[31, 32] (DFT) calculations and Wannier functions[33-35], we explore the band crossings throughout the full Brillouin zone (BZ) between the nominally highest valence and lowest conduction bands in $\gamma$-$PtBi_2$ with two different experimental structural parameters and the averaged one. The main difference between the two experimentally reported structural parameters[1, 12] is the magnitude of the Bi-layer buckling. In contrast to the robustness of WNL on the $\Gamma$-$K$-$A$ plane, the bulk gap region and the Set 1 Weyl points (WPs) along the $\bar{\Gamma}$-$\bar{M}$ direction, we find the number and locations of the extra sets of WPs are sensitive to the Bi-layer buckling. The averaged experimental structure gives more extra sets of WPs. We also study the evolution of the 2D Fermi surface (FS) with FAs and surface states around the bulk gap region. Although the scattering vectors of the calculated quasi-particle interference (QPI) near $E_F$ are similar for different Bi-layer buckling and in good agreement with ARPES[10, 16] and experimental

QPI[17, 20], we find variations and secondary features of FAs and QPI with different Bi-layer buckling, including the additional outward FAs on the other side of WP projection at higher energy. Our detailed mapping of the WNL, multiple WP sets, FAs and QPI in $\gamma$-$PtBi_2$ provides a better understanding of experimental observations.

## II. Computational Methods

Band structures of $\gamma$-$PtBi_2$ in space group 157 with spin-orbit coupling (SOC) have been calculated in DFT with PBE[36] exchange-correlation (XC) functional and modified Becke-Johnson (mBJ)[37] exchange potential, a plane-wave basis set and projected augmented wave method[38] as implemented in VASP[39, 40]. A Monkhorst-Pack[41] (8×8×8) $k$-point mesh with a Gaussian smearing of 0.05 eV including the $\Gamma$ point and a kinetic energy cutoff of 230.3 eV have been used. Besides the two experimental[1, 12] and averaged structural parameters for band structure calculations, the atomic positions have also been relaxed in PBE[36] and XC functionals with van der Waals (vdW) interactions including D3[42], optB86b[43], r2SCAN+rVV10[44] with lattice constants fixed to experimental values[1] to compare the relaxed internal structural parameters to experiment. The atomic positions are relaxed with the remaining absolute force on each atom being less than 0.02 eV/Å. To explore the band structure and surface states, tight-binding models based on maximally localized Wannier functions[33-35] have been constructed to reproduce closely the bulk band structure including SOC in the range of $E_F$±2eV with Pt *sd* and Bi *p* orbitals. The surface spectral function, 2D Fermi surface with Fermi arcs of the semi-infinite surface and QPI as joint density of states[45] have been calculated with the surface Green's function methods[46, 47] as implemented in WannierTools[48]. For the topological band structure calculations with the three structural parameter sets, their different number of WPs and momentum-energy have been checked with a higher kinetic energy cutoff of 400.0 eV and denser (12×12×12) $k$-point mesh for convergency.

## III. Results and Discussion

### III-A. Weyl nodal loop and Weyl points

Figure 1 presents the bulk crystal and band structures of trigonal $\gamma$-$PtBi_2$. The crystal structure of $\gamma$-$PtBi_2$ in space group P31m (157) is shown in Fig.1(a), which has no inversion symmetry, but a set of vertical mirror planes combined with the 3-fold rotation symmetry. The structure is formed with stacking Bi-Pt-Bi layers along the *c*-axis, which has three different types of Bi sites, giving two different terminations on (0001) surface. The A-termination has a buckled Bi-layer with Bi1 and Bi2 sites, while the B-termination has a flat Bi-layer with Bi3 site. These two different terminations have shown different SC behavior and surface band structures. For the structural parameters of $\gamma$-$PtBi_2$, there are two experimental reports, one in 2014[1] and the other in 2020[12], which has a major difference in the magnitude of the Bi-layer buckling in the A-termination. As listed in Table 1, between the two sets of experimental structural parameters for $\gamma$-$PtBi_2$, the lattice constants of *a* and *c* are almost the same, which are also plotted in Fig.2 for the deviation with respect to the 2014 experimental values[1]. The internal coordinates of Pt and Bi3 sites, i.e., $x_{Pt}$, $x_{Bi3}$ and $z_{Bi3}$ are also similar between the two experiments as shown by the red bars in Fig.2. However, there is a sizable difference for the *z* values of Bi2 and Bi1 sites as their in-plane positions are at the ideal (0, 0) and (1/3, 2/3). The 0.292 of $z_{Bi2}$ in Ref. [12] is larger than the 0.28429 in Ref. [1], while the 0.137 of $z_{Bi1}$ is smaller than the 0.15486, which results in a larger Bi-layer buckling for the vertical interlayer distance between Bi2 and Bi1, $d_{Bi2\text{-}Bi1}$, of 0.955 Å in Ref. [12] than 0.798 Å in Ref. [1]. Such a difference of 0.157 Å as plotted in Fig.2 seems small at first glance, but we will show later that they result in different sets of WPs as well as some difference in FA features.

To validate the difference in the experimental structural parameters, we have used four different types of XC functionals in DFT to relax the internal atomic positions by fixing the unit cell lattice constants to Ref.[1], because the difference in the two experimental lattice constants is very small. The four XC functionals include PBE and three with explicit vdW interactions, namely, D3 with empirical vdW interaction[42], optB86b with nonlocal vdW functional[43] and the latest r2SCAN+rVV10[44]. Among the four relaxed structures as listed in Table 1 and plotted in Fig.2, again, the deviations of Pt and Bi3 are small, and their values are close to those reported in both experiments. For $z_{Bi2}$, the four relaxed values are all smaller than Ref.[1]. For $z_{Bi1}$, the relaxed values are in between the two experimental values and closer to Ref.[1], except for r2SCAN+rVV10, which is slightly larger than Ref.[1]. As for the Bi-layer buckling of $d_{Bi2\text{-}Bi1}$, the relaxed values are slightly larger except for r2SCAN+rVV10 with a slightly smaller value than Ref.[1]. Overall, as compared and plotted in Fig.2, the DFT-relaxed atomic positions are closer to those of Ref.[1] than Ref.[12]. Usually using experimental lattice constants to correct the over/underestimation of DFT with relaxed atomic positions are good choices for band structure calculations. Here the two reported lattice constants are almost the same, while the largest difference is in $z_{Bi2}$ and $z_{Bi1}$, i.e., with more or less Bi-layer buckling. Thus, we will also include the averaged structural parameters between these two experimental reports as listed in Table 1 and plotted in Fig.2 (blue bar) to study the changes in band structure induced by the different magnitude of the Bi-layer buckling.

The projected density of states (PDOS) in Fig.1(b) shows that the band hybridization and chemical bonding are mostly between the Pt 5*d* and Bi 6*p* orbitals. Around the $E_F$, there is a local minimum of DOS. When inspecting the band structure in Fig.1(d), along the *Γ-M-K-Γ* direction on the $k_z$=0.0 plane, it is semi-metallic with a major band inversion along the *Γ-M* direction. But going to the $k_z$=0.5 plane, the band structure is more metallic with large dispersive band crossing $E_F$. There are totally five bands crossing the $E_F$ and the 3D Fermi surface is plotted in Fig.1(c). The major FS sheets are near the $k_z$=0.5, while near the $k_z$=0.0 plane, there are pairs of small electron and hole pockets, echoing the band structure plot in Fig.1(d). The touching pockets near $k_z$=0.0 plane gives the robust Set 1 WPs from the main band inversion along the *Γ-M* direction, while the large pockets near $k_z$=0.5 gives the WNL and most of the other sets of WPs. For the WNL, as zoomed in Fig.1(f) along the *H-A* direction, there are two crossing points (green arrows), one at $E_F$ and the other near $E_F$–0.80 eV, showing the large range of dispersion for these bands. To confirm the symmetry protection of WNL by the $M_x$ mirror plane, we plot the $M_x$ mirror eigenvalue and overlaid on top of the band dispersion in Fig.1(g) along the *H-A* direction, verifying the opposite $M_x$ mirror eigenvalues of the WNL crossings.

By fitting to a tight-binding model with Wannier functions to reproduce the band structure closely within $E_F$±2.00 eV, we can efficiently search over the full BZ to map out the WNL as plotted in green in Fig.1(e) with the contours of zero band gap. Note these WNL are closed loops with periodic boundary condition along the $k_z$ direction. With the averaged structural parameters, we also find six sets of WPs which are plotted in Fig.1(e) together with the WNL, and their momentum-energy are listed in Table 2. Each listed WP can be used to produce the other 11 WPs in the same set because they are related by time-reversal, rotation and mirror symmetries with the corresponding chirality. To better show WP positions, we have projected them on the $k_x$-$k_y$ or (0001) plane in the surface BZ (SBZ) as shown in Fig.1(h). Among them, Set 1 WPs are halfway along the $\bar{\Gamma}$-$\bar{M}$ direction as labeled, while the other five sets are near the $\bar{\Gamma}$-$\bar{K}$ direction and close to the WNL. The band dispersion of WP set 1 and 2 are plotted in Fig.1(i) as they are near the $E_F$. Fig.(j) and (k) plot the band dispersions of the other four sets of WPs, which are further above the $E_F$.

When the input structural parameters are switched to Ref.[12] with more Bi-buckling, the number of WP set is reduced to two as listed in Table 2. Set 1 WPs remain just above the $E_F$ at 0.033 eV and still close to the $\Gamma$-$M$ direction. In contrast, when switching to Ref.[1] with less Bi-buckling, the number of WP sets becomes five as listed in Table 2. Besides Set 1 WPs being moved to 0.072 eV, a new set of WPs (Set 2) appear just below the $E_F$ at −0.015 eV and is also close to the $\Gamma$-$M$ direction and $k_z$=0.0 plane. With all the three structural parameters, the WNL is persistent because of the protection from the mirror symmetry and Set 1 WPs are also robust because it is from the main band inversion along the $\Gamma$-$M$ direction. The WNL using the three structural parameters with different Bi-layer buckling are plotted in Fig.2(b-d), respectively, as the contours of zero band gap on the $k_y$-$k_z$ plane. The shape of the WNL remains the same with only main difference in the range along the $k_z$ direction near $k_y$=0.3 (1/Å). The robustness of WNL is due to the same symmetry protection from the $M_x$ mirror plane.

The extra set of WPs about 0.16 eV above the $E_F$ has been noticed before with the more Bi-layer buckling structure[15]. But here we find it is remarkable that the two different experimental structural parameters give quite different number of extra WP sets, although the difference in Bi-layer buckling is only 0.157 Å. Importantly from this structural variation, we demonstrate that Set 1 WPs just above the $E_F$ and close to the $\Gamma$-$M$ direction are very robust, while the extra sets of WPs at higher energy and close to the $\Gamma$-$K$ direction are more susceptible to the change in structural parameters. In a way, the WNL acts as a generator of extra sets of WPs with additional band inversion just off the $\Gamma$-$K$ direction, which does not necessarily have a linear behavior regrading to the Bi-layer buckling as shown by the different numbers of WP sets. In the remaining of the paper, we will present and compare the surface band structures from these three different structural parameters.

## III-B. Fermi arcs, quasi-particle interference and scattering vectors

Figures 3 and 4 compare the 2D FS with FAs and QPI of $\gamma$-$PtBi_2$ (0001) calculated with the three different structural parameters, i.e., the averaged structure, the experimental structure with more Bi-layer buckling from Ref.[12] and less Bi-layer buckling from Ref.[1], around the $E_F$ at −0.012 eV and +0.012 eV, respectively. The 2D FS from Fig.3 and 4 are also zoomed in Fig.5 around Set 1 WPs. Foremost, all the three cases have the projected bulk gap region around 2/3 along the $\bar{\Gamma}$-$\bar{M}$ direction, which is the signature from the main band inversion along the $\Gamma$-$M$ and $A$-$L$ directions as seen in Fig.1(d). Figure 3 also has arrows highlighting the scattering vectors in both 2D FS and QPI for better comparison among them and also to experimental ARPES and QPI. In Fig.3(a) for the 2D FS at −0.012 eV for the A-termination with the averaged structure, the FAs compare well with ARPES (Fig.3(a) of Ref.[16]) showing a waterdrop shape, and the mergence into the projection of Set 1WPs is better seen in Fig.4(a) at +0.012 eV with zoomed in Fig.5(a) and Fig.5(c). The scattering vectors transforming FAs to QPI patterns are presented in Fig.3(a) and (b). From the short to long scattering vectors in Fig.3(a), $q_1$ is the nearest to connect FAs between neighboring SBZ across the $\bar{M}$ point, while $q_2$ is the nearest to connect FAs within the same SBZ. Then $q_3$ and $q_4$ are the next nearest to connect FAs between neighboring SBZ and within the same SBZ, respectively. For QPI in Fig.3(b), these scattering vectors are shifted to the origin of the SBZ and give the QPI patterns in two groups. The $q_1$ and $q_2$ are both along the $\bar{\Gamma}$-$\bar{M}$ direction, giving two-pedal shapes as highlighted by the dashed oval. The $q_3$ and $q_4$ are around the $\bar{K}$ point giving three-leaf shapes as highlighted by the dashed triangle.

The similar FAs and QPI are also seen from the calculations with the more Bi-layer buckling structure in Fig.3(e) and (f) with the zoomed FAs in Fig.5(e). The scattering vectors from $q_1$ to $q_4$ remain the same when comparing Fig.3(e-f) for more Bi-layer buckling with Fig.3(a-b) for the averaged structure, because the bulk gap region and similarly shaped FAs are in the same location, i.e., 2/3 along the $\bar{\Gamma}$-$\bar{M}$ direction. The more roundish shape of the FAs for more Bi-layer buckling than the averaged structure is reflected in the more roundish QPI pattern in Fig.3(f) for both the two-pedal in the dashed oval and the three-leaf in the dashed triangle. The QPI in Fig.3(f) also agrees well with the previous calculation (Fig.3(C) of Ref.[17]). We notice that with the same more Bi-layer buckling structure[12], we can only get the two sets of WPs and the associated FAs using mBJ, but not the regular PBE XC functional in DFT as specified in Ref.[15].

In contrast, the 2D FS in Fig.3(i) from the less Bi-buckling structure has two sets of WP projections along the $\bar{\Gamma}$-$\bar{M}$ direction, which give a different shape of FAs, smaller and connecting on the other side of Set 2 WP as zoomed in Fig.5(i). But importantly, these FAs are still bounded similarly by the bulk gap region, 2/3 along the $\bar{\Gamma}$-$\bar{M}$ direction. Thus, the scattering vectors from $q_1$ to $q_4$ still apply, but with some adjustment. As seen in Fig.3(j), the $q_1$ and $q_2$ along the $\bar{\Gamma}$-$\bar{M}$ direction (dashed oval) become more separated. Similarly, the $q_3$ and $q_4$ around the $\bar{K}$ point (dashed triangle) are also more spread out. In addition, the secondary FAs at 2/3 along the $\bar{\Gamma}$-$\bar{K}$ direction (unlike the primary FAs at the bulk gap region 2/3 along the $\bar{\Gamma}$-$\bar{M}$ direction) can be connected by $q_5$ in Fig.3(i) and become visible because of the reduced primary FAs with the less Bi-layer buckling. Thus, although the QPI pattern in Fig.3(j) for less Bi-layer buckling looks different in first glance from those in Fig.3(f) and (b), they are still related variants with the same sets of scattering vectors (from $q_1$ to $q_4$) with adjustment, plus an additional secondary vector of $q_5$. The QPI pattern in Fig.5(j) is different from previous experimental QPI[17], but is more in line with the recent experimental QPI measured at the quasiparticle edge of the SC gap (Fig.3(a) of Ref.[20]) with more separated $q_3$ and $q_4$ around the $\bar{K}$ point as well as the weak secondary $q_5$. It is possible to observe different QPI patterns from different samples, because there are two different experimentally reported structural parameters with different Bi-layer buckling, which may be due to small off-stoichiometry resulted from different growth condition as well as also affected by strain.

Moving to the B-termination in the bottom half of Fig.3, all three structures show similar FAs around the far side of the bulk gap region, although for the less Bi-buckling structure, this FA is spanned by Set 2 WPs just below the $E_F$ as also zoomed in Fig.5(j). The simulated QPI patterns for the B-termination in the last row of Fig.3 for the three structures agree with each other overall and also the previous experimental QPI (Fig.3(D) of Ref.[17]). The correspondence among the same sets of scattering vectors of $q_1$ and $q_2$ in dashed oval, and $q_3$ and $q_4$ in dashed triangle are more clearly seen here for B-termination than the A-termination, although the relative intensity among different region can vary differently. Among them, Fig.3(l) from the less Bi-layer buckling structure gives the clearest QPI pattern. These above analyses emphasize the close relation and range of variation for FAs and QPI that can happen in $\gamma$-$PtBi_2$ upon the structural modulation of Bi-layer buckling as seen from the two different experimentally reported structural parameters.

In Fig.4 at +0.012 eV, while the FAs for the averaged structure in Fig.4(a) and more Bi-buckling structure in Fig.4(e) are merged to the rim of the bulk gap region as zoomed in Fig.5.(c) and (g), respectively, the less Bi-layer buckling structure in Fig.4(i) has the FAs switch toward Set 1 WPs projection with the far side half of the ring breaking off due to Set 2 WPs as zoomed in Fig.5(k). Its calculated QPI in Fig.4(j) now overall resembles more to the other two structures in Fig.4(b) and (f), but with some local features like the star pattern at the $\bar{K}$ point. On the B-termination, the FAs from the far side of bulk gap region in Fig.4(c) and (g) show reduced intensity, except for the less Bi-layer buckling in Fig.4(k), whose FAs intensity is higher as they are spanned by Set 2 WPs with zoomed in Fig.5(l). Interestingly for all the three structures, the 2D FS show additional FAs pointing outward for Set 1 WPs (see $k_x$=0.3 (1/Å) in Fig.5(d), (h) and (l)). These line-shaped FAs are aligned along the inner hexagon of bulk band projection around the $\bar{\Gamma}$ point, which have not been found before in earlier studies and are the new features we predict for the (0001) surface of $\gamma$-$PtBi_2$. In our early study[49] of $ZrTe_5$, it has been shown that the inward or outward FA connection can switch direction with different lattice distortions as induced by coherent phonons. Here for $\gamma$-$PtBi_2$, the inversion symmetry is already broken, while the switching of the FA connection happens at different binding energies.

Despite the differences in the number of WP sets, the bulk gap region 2/3 along the $\bar{\Gamma}$-$\bar{M}$ direction is the common theme for all the three structures. To better understand the dispersion of the FA associated surface states, using the averaged structure, we plot in Fig.6 the surface spectral functions with cuts indicated in Fig.6(a) with the bulk only contribution in the last column. The 2D FS at $E_F$ in Fig.6(a) and (e) for the A and B-terminations have the similar FA features to the +/−0.012 eV in Fig.3 and 4. For the first cut passing directly through a pair of Set 1 WPs, the $W_1^-$ and $W_1^+$ at $E_F$+0.055 eV can be clearly seen in Fig.6(j) as labeled near the bulk gap region with vertical lines. For the A-termination in Fig.6(b), the surface states follow the bottom rim of the bulk gap region and converge to Set 1 WP projection. On the B-termination in Fig.6(f), the surface states are instead overlayed on the bulk band projection but also converge to Set 1WP projection. Moving to the second cut on the right in Fig.6(k), the top of bulk valence band in the middle goes up in energy to +0.130 eV. The surface states for B-termination in Fig.6(g) also go up in energy. Interestingly for A-termination in Fig.6(c), the surface states are still at the bottom rim of the bulk gap region, showing the robustness. With the third cut further to the right in Fig.6(l), the top of bulk valence bands in the middle are pushed up even higher in energy to +0.200 eV and cross with the bulk conduction bands to have more accidental crossings as confirmed by the extra sets of WPs at +0.200 eV and above. At the same time, the bottom of bulk conduction bands on the other side of the bulk gap region in Fig.6(l) drops to lower energy and almost touches the bulk valence band around the $E_F$, which can result in the Set 2 WPs just below the $E_F$ as seen for the less Bi-buckling structure. For A-termination in Fig.6(d), the surface states at the bottom rim of bulk gap region remain robust. Interestingly for the B-termination in Fig.6(h), the original surface states are pushed to energy above 0.120 eV, but new surface states emerge at the bottom rim of the bulk gap region, giving the FAs at the far side of the bulk gap region in 2D FS of Fig.6(e).

As mentioned earlier, the FAs connection between the robust Set 1 WP can switch direction at higher energy. Also to explore the FAs for the extra sets of WPs, we have plotted 2D FS and QPI in Fig.7 at +0.100 eV for the averaged structure. For the A-termination in Fig.7(a), the surface states form the FAs connecting among the extra sets of WPs as more clearly seen in the surface only contribution in Fig.7(c). For the B-termination in Fig.7(b), the FAs are the curly surface states (Fig.7(d)) stemming outward from Set 1 WPs with high intensity. Correspondingly, the QPI for A-termination in Fig.7(e) shows a different pattern at the higher energy from the QPI near $E_F$ in Fig.3 and 4. The QPI for B-termination in Fig.7(f) has more prominent features from the curly FAs. These calculated QPI patterns at $E_F$+0.100 eV are from the predicted new FA features at higher energy and need future experiments to investigate.

## IV. Conclusion

In conclusion, we have used first-principles calculations and Wannier functions to explore the band crossings throughout the full Brillouin zone (BZ) between the nominally highest valence and lowest conduction bands in $\gamma$-$PtBi_2$ with different structural parameters. We find that the main difference between the two experimentally reported structural parameters[1, 12] is the magnitude of the Bi-layer buckling. By calculating band structures of both more and less Bi-layer buckling structures as well as the averaged one, we study the Weyl nodal loop (WNL) and multiple sets of Weyl points (WPs). The robust features are the WNL on the $\Gamma$-$K$-$A$ plane as protected by the mirror symmetry, the bulk gap region, and the set of WPs just above the Fermi energy ($E_F$) from the main band inversion along the $\bar{\Gamma}$-$\bar{M}$ direction. In contrast, the number and locations of other multiple sets of WPs are sensitive to the Bi-layer buckling. We study the evolution of the Fermi arcs around the bulk gap region. Besides calculating the 2D Fermi surface and quasi-particle interference (QPI) in good agreements with ARPES and experimental QPI, we predict new Fermi arc features and the related QPI pattern at $E_F$+100 meV, which need to be verified in future experiments.

**Data Availability**: The data that support the findings of this study are openly available[50].

## Acknowledgements

We acknowledge helpful discussions with Adam Kaminski, Hermann Suderow and Paul C. Canfield. Work on bulk and surface band structures was supported by the U.S. Department of Energy (DOE), Office of Science, Basic Energy Science, Materials Science & Engineering Division through the Ames National Laboratory. Work on topological and symmetry analysis was supported by the Center for the Advancement of Topological Semimetals, an Energy Frontier Research Center funded by U.S. Department of Energy (DOE), Office of Science, Basic Energy Sciences. The Ames National Laboratory is operated for the U.S. Department of Energy by Iowa State University under Contract No. DE-AC02-07CH11358. Some of this research used resources of the National Energy Research Scientific Computing Center (NERSC), a DOE Office of Science User Facility.

Table 1. Experimental[1, 12], averaged and relaxed lattice parameters of trigonal $\gamma$-$PtBi_2$. For ionic relaxation, the unit cell lattice constants are fixed to the experimental values with the internal coordinates relaxed in different exchange-correlation functionals of PBE[36], D3[42], optB86b[43] and r2SCAN+rVV10[44].

| | $a$ (Å) | $c$ (Å) | $x_{Pt}$ | $x_{Bi3}$ | $z_{Bi3}$ | $z_{Bi2}$ | $z_{Bi1}$ | $d_{Bi2-Bi1}$ (Å) |
|---|---|---|---|---|---|---|---|---|
| Ref.[12] | 6.5731 | 6.1619 | 0.2619 | 0.3856 | 0.767 | 0.292 | 0.137 | 0.955 |
| Ref.[1] | 6.573 | 6.1665 | 0.263 | 0.38876 | 0.77284 | 0.28429 | 0.15486 | 0.798 |
| Avg. | 6.57305 | 6.1642 | 0.26245 | 0.38718 | 0.76992 | 0.28814 | 0.14593 | 0.877 |
| PBE | | | 0.26030 | 0.39193 | 0.77795 | 0.28014 | 0.14694 | 0.821 |
| D3 | | | 0.26310 | 0.39018 | 0.77897 | 0.27885 | 0.14870 | 0.803 |
| optB86b | | | 0.26067 | 0.39108 | 0.77803 | 0.27986 | 0.14730 | 0.817 |
| r2SCAN +rVV10 | | | 0.26506 | 0.38716 | 0.77447 | 0.28081 | 0.15645 | 0.767 |

Table 2. Momentum-energy of different sets of Weyl point (WP) using the experimental[1, 12], and averaged lattice parameters of trigonal $\gamma$-$PtBi_2$. The band structures are calculated with mBJ and fitted to a tight-binding Hamiltonian with Wannier functions.

| Lattice | WP set | Momentum ($k_x$,$k_y$,$k_z$) (1/Å) | E-$E_F$ (eV) |
|---|---|---|---|
| Avg. | 1 | (0.33216, 0.03110, –0.13023) | 0.055 |
| | 2 | (0.43218, 0.14289, –0.27762) | 0.234 |
| | 3 | (0.37573, 0.13282, –0.29144) | 0.279 |
| | 4 | (0.40425, 0.19143, –0.37676) | 0.313 |
| | 5 | (0.40396, 0.14193, –0.30754) | 0.325 |
| | 6 | (0.39015, 0.19729, –0.35641) | 0.339 |
| Ref.[12] | 1 | (0.31898, 0.03920, –0.14161) | 0.033 |
| | 2 | (0.34309, 0.11210, –0.26517) | 0.181 |
| Ref.[1] | 1 | (0.34067, 0.02352, –0.11875) | 0.072 |
| | 2 | (0.43178, 0.01384, –0.03291) | –0.015 |
| | 3 | (0.45107, 0.13725, –0.28224) | 0.226 |
| | 4 | (0.44785, 0.13626, –0.36634) | 0.319 |
| | 5 | (0.39474, 0.16978, –0.32786) | 0.356 |

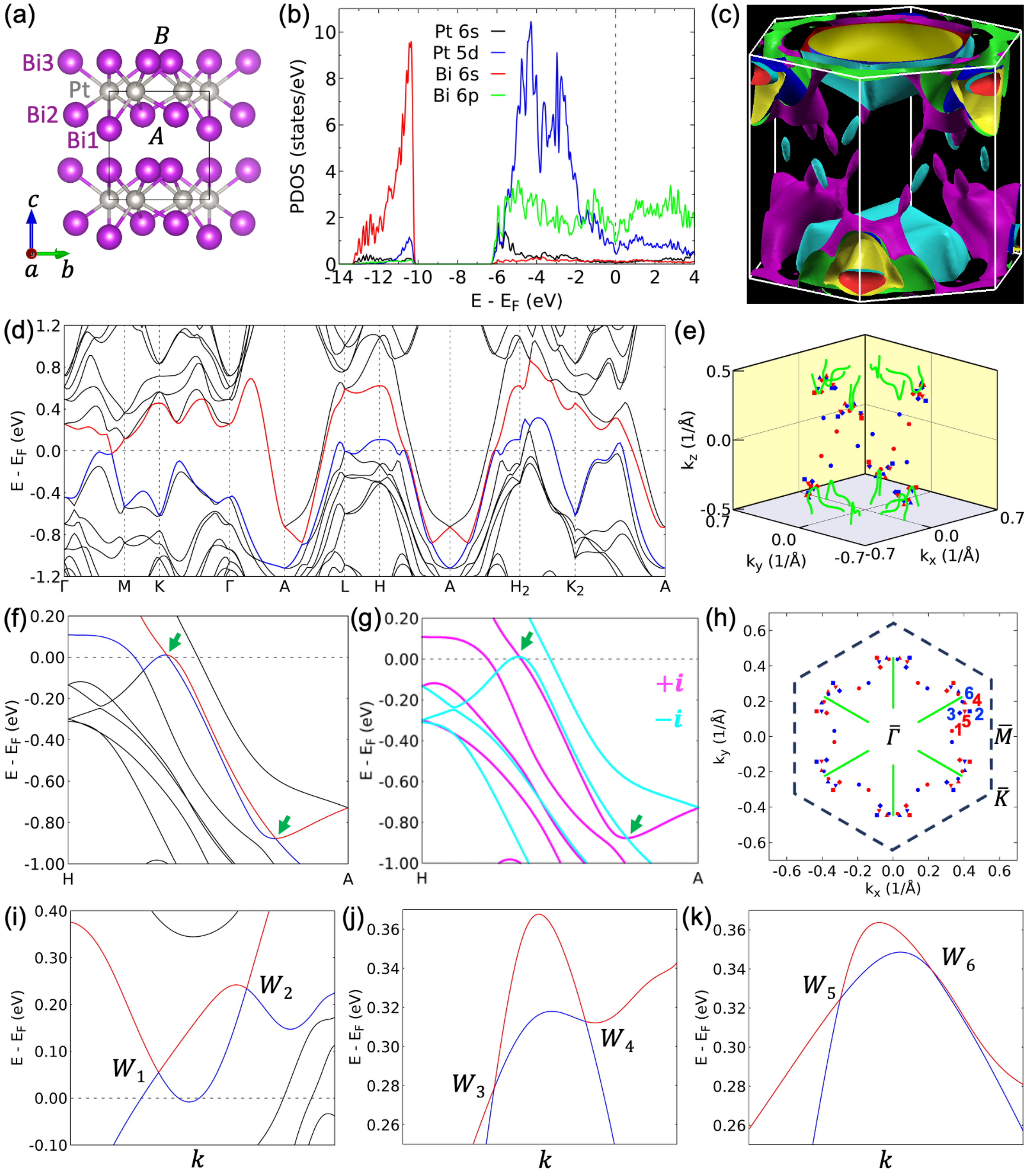

Figure 1. Band structure of $\gamma$-$PtBi_2$. (a) Crystal structure with (0001) surface A- and B-terminations. (b) Projected density of states (PDOS), (c) 3D Fermi surface, (d) band structure along high-symmetry directions. The nominally highest valence and lowest conduction bands are in blue and red, respectively. (e) Six sets of Weyl points (WPs) in red/blue for opposite chirality and one set of Weyl nodal loop (WNL) in green. WNL are closed loops with periodic boundary condition along $k_z$. (f) Band structure zoomed along the *H*-*A* direction and (g) overlaid with $M_x$ mirror eigenvalues ($\pm i$) with green arrows pointing to the WNL crossings. (h) Projection on the (0001) plane of (e) with the six sets of WPs labeled together with WNL. (i-k) Band dispersion near the six set of WPs as labeled.

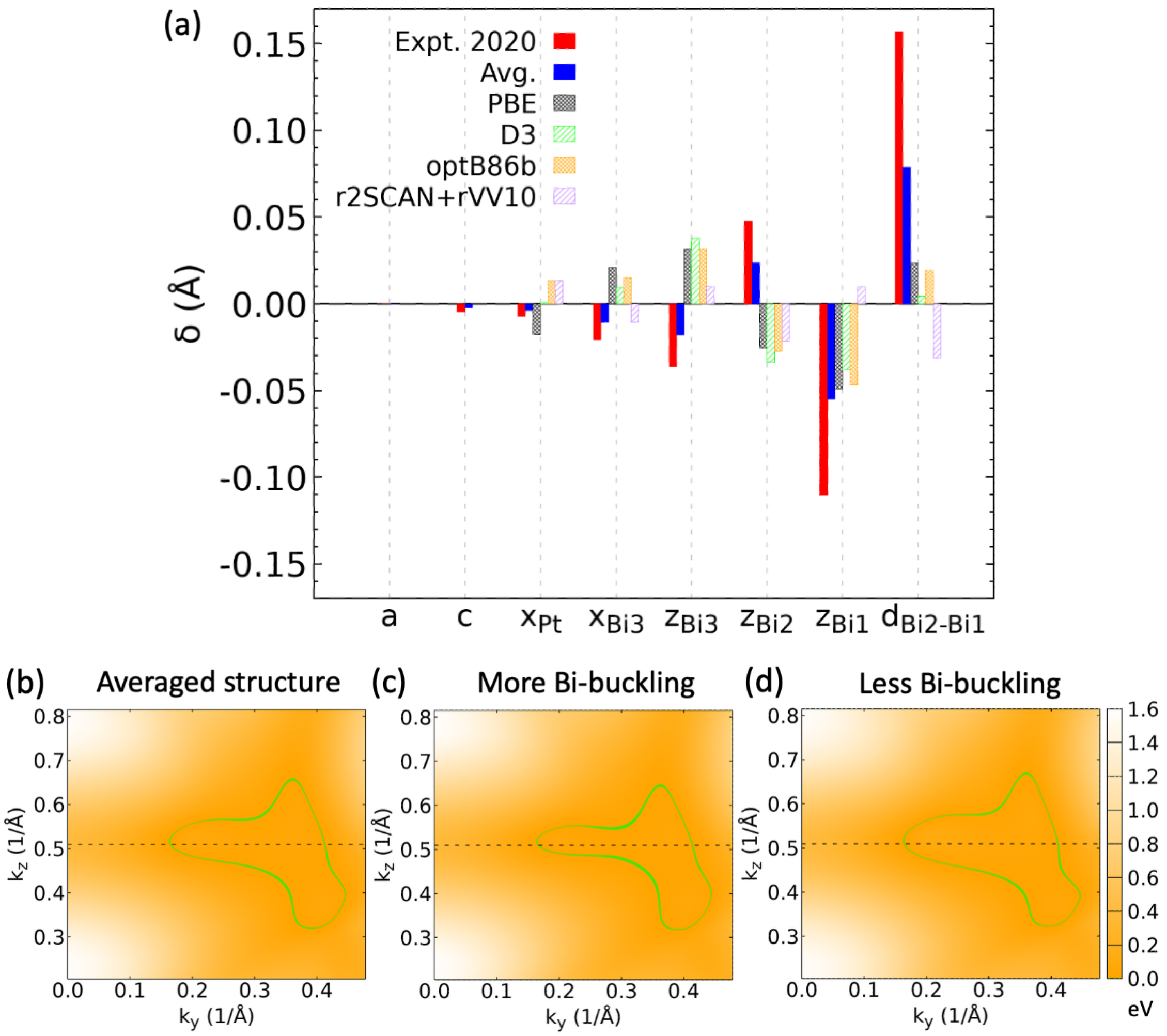

Figure 2. Difference of structural parameters of $\gamma$-$PtBi_2$ with respect to the experimental values in 2014 of Ref.[1]. The corresponding data of these parameters are listed in Table 1. The ionic relaxations in DFT are with exchange-correlation functionals of PBE, D3, optB86b and r2SCAN+rVV10, respectively. The averaged (Avg.) data are between the two experimental data of Ref.[1] in 2014 and Ref.[12] in 2020. Zero-gap contour showing the Weyl nodal loop (WNL) in green on the $k_y$-$k_z$ or $M_x$ mirror plane for (b) the averaged structure, experimental structure with (c) more Bi-buckling in Ref.[12], and (d) less Bi-buckling in Ref.[1], respectively. The intensity of orange color shows the band gap size in eV as indicated by the color bar on the right.

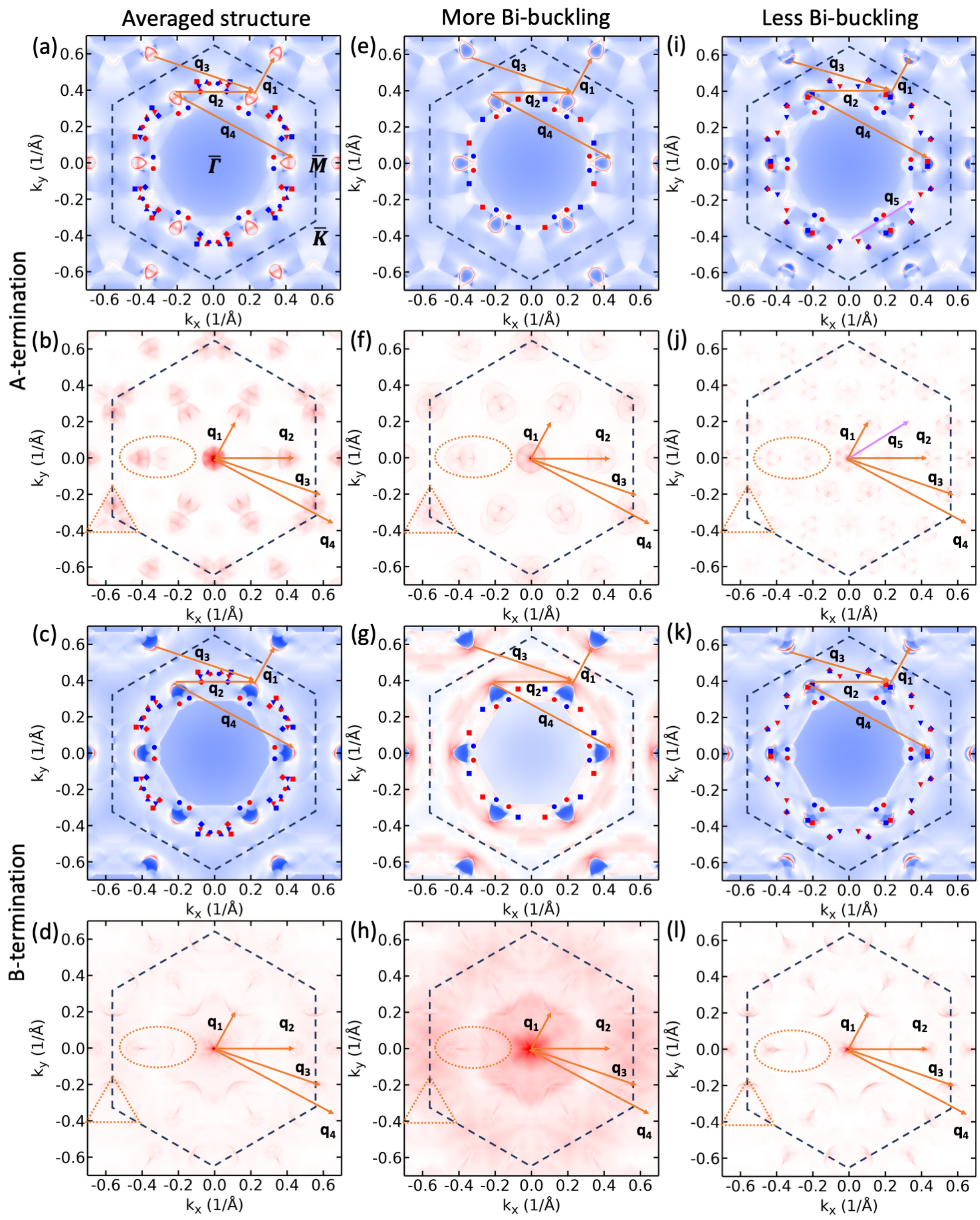

Figure 3. 2D Fermi surface and QPI of $\gamma$-$PtBi_2$ (0001) at $E_F$−0.012 eV calculated with different structural parameters. Panel (a-d) are for the averaged structure, (e-h) for Ref.[12] with more Bi-buckling and (i-l) for Ref.[1] with less Bi-buckling. The top (bottom) two rows are for A (B) termination with 2D Fermi surface showing Fermi arcs (FAs) and the corresponding QPI. For Fermi surface, red (blue) stands for high (low) intensity. For QPI, red (white) stands for high (low) intensity. The multiple sets of Weyl points are labeled in different point shapes with red (blue) for positive (negative) chirality. Orange and purple arrows are the scattering vectors among FAs (from $q_1$ to $q_5$) and the resulting QPI vectors. The oval (triangle) is to highlight the associated features of $q_1$ and $q_2$ ($q_3$ and $q_4$).

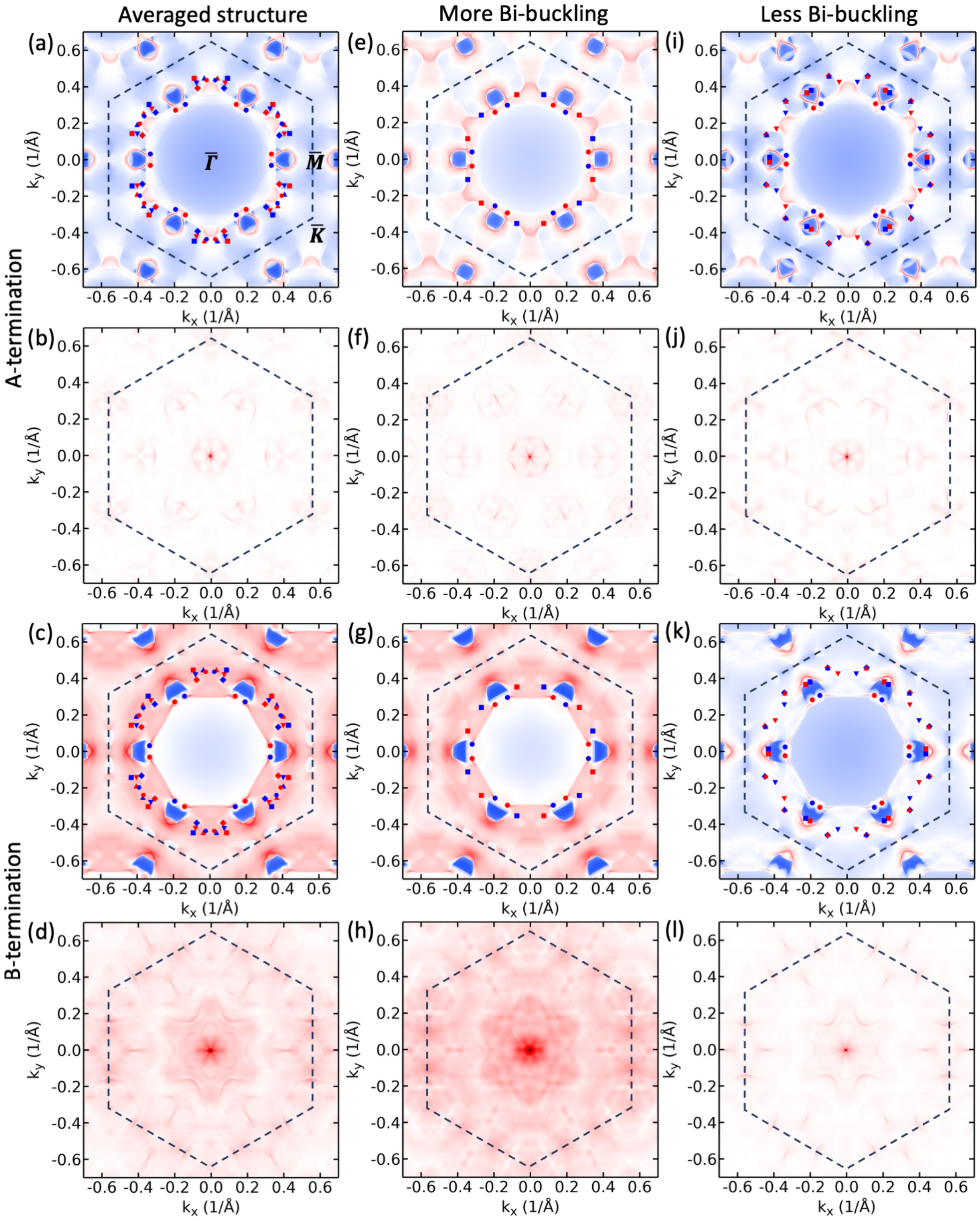

Figure 4. 2D Fermi surface and QPI of $\gamma$-$PtBi_2$ (0001) at $E_F$+0.012 eV calculated with different structural parameters. Panel (a-d) are for the averaged structure, (e-h) for Ref.[12] with more Bi-buckling and (i-l) for Ref.[1] with less Bi-buckling. The top (bottom) two rows are for A (B) termination with 2D Fermi surface showing Fermi arcs (FAs) and the corresponding QPI. For Fermi surface, red (blue) stands for high (low) intensity. For QPI, red (white) stands for high (low) intensity. The multiple sets of Weyl points are labeled in different point shapes with red (blue) for positive (negative) chirality.

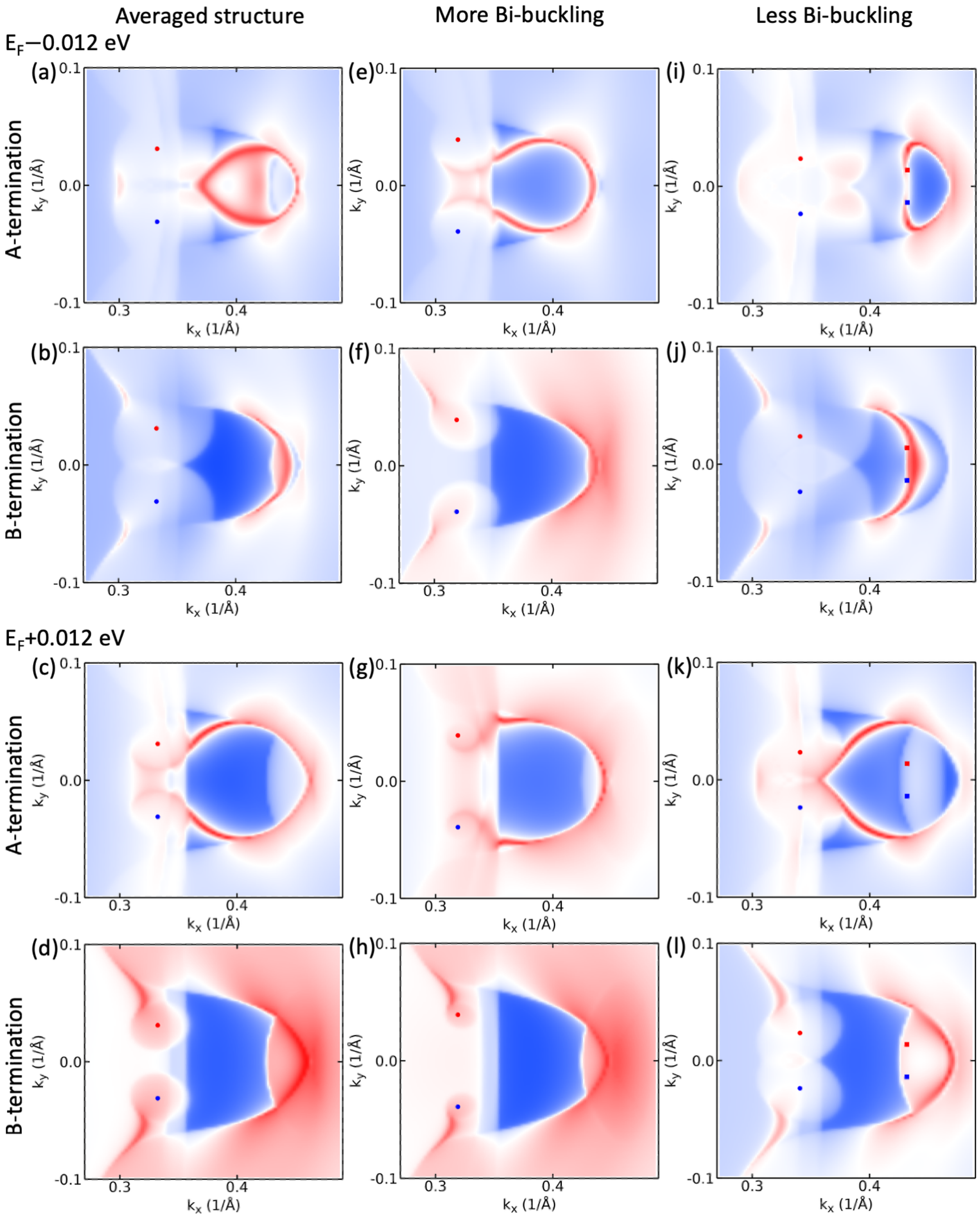

Figure 5. 2D Fermi surface of $\gamma$-$PtBi_2$ (0001) zoomed in around the projection of Set 1 Weyl point calculated with different structural parameters. Panel (a-d) are for the averaged structure, (e-h) for Ref.[12] with more Bi-buckling and (i-l) for Ref.[1] with less Bi-buckling. For each column, top (bottom) two rows are for $E_F$–0.012 eV ($E_F$+0.012 eV) with A and B-terminations. For Fermi surface, red (blue) stands for high (low) intensity. For QPI, red (white) stands for high (low) intensity. The associated sets of Weyl points are labeled in 2D Fermi surface as different point shapes with red (blue) for positive (negative) chirality.

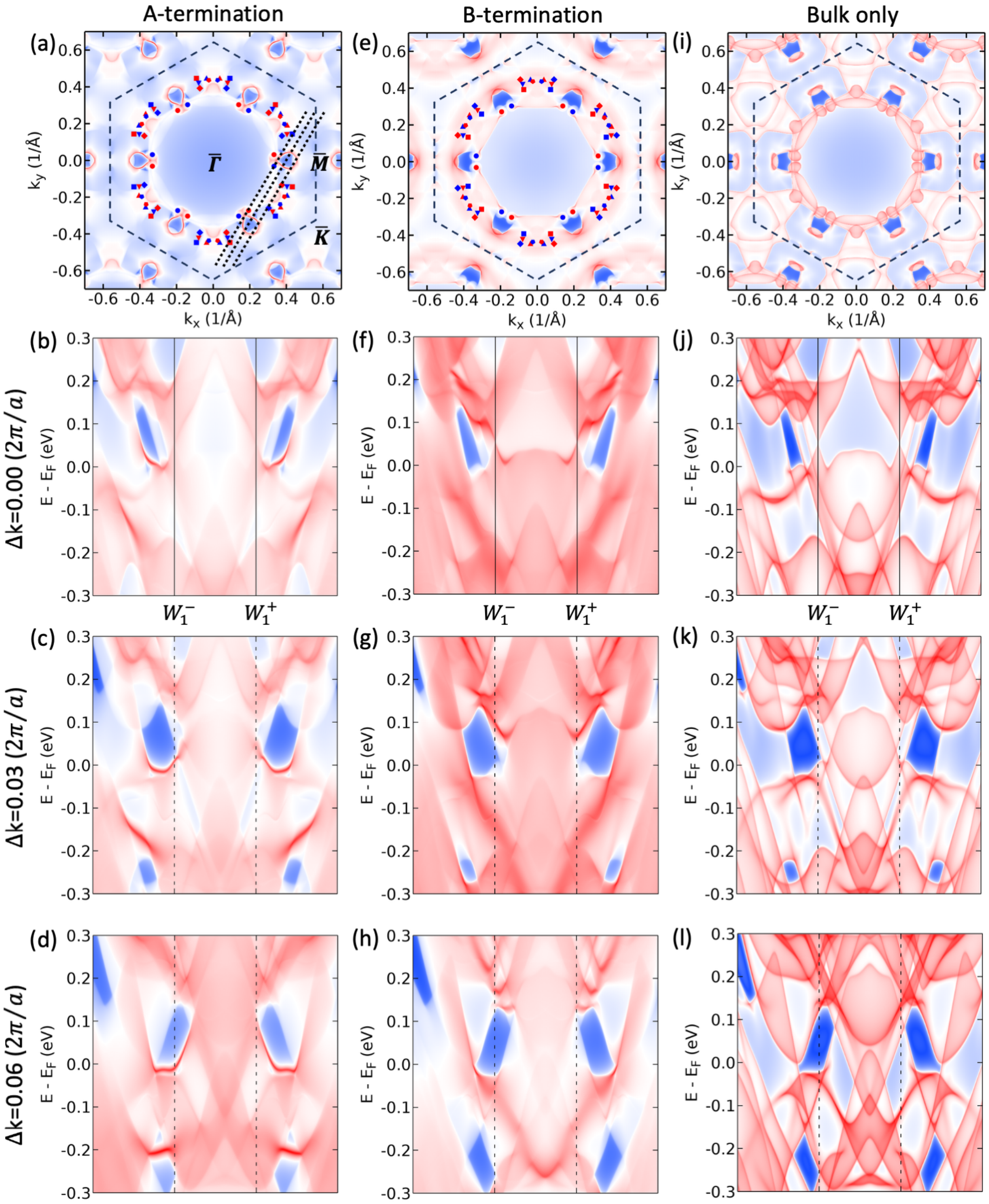

Figure 6. Surface spectral function of $\gamma$-$PtBi_2$ (0001) calculated with the averaged experimental structural parameters. The columns are for A, B-termination and bulk only contributions. Panel (a-b) are the 2D Fermi surface with the multiple sets of Weyl points labeled as different symbol shapes with red (blue) for positive (negative) chirality. In (a), the first dashed line on the left is the cut across the Set 1 Weyl points (W1) with the surface spectral function shown in the rows. The two additional dashed lines are cuts shifted along the b1-axis as labeled in reach row. In each panel, red (blue) stands for high (low) intensity.

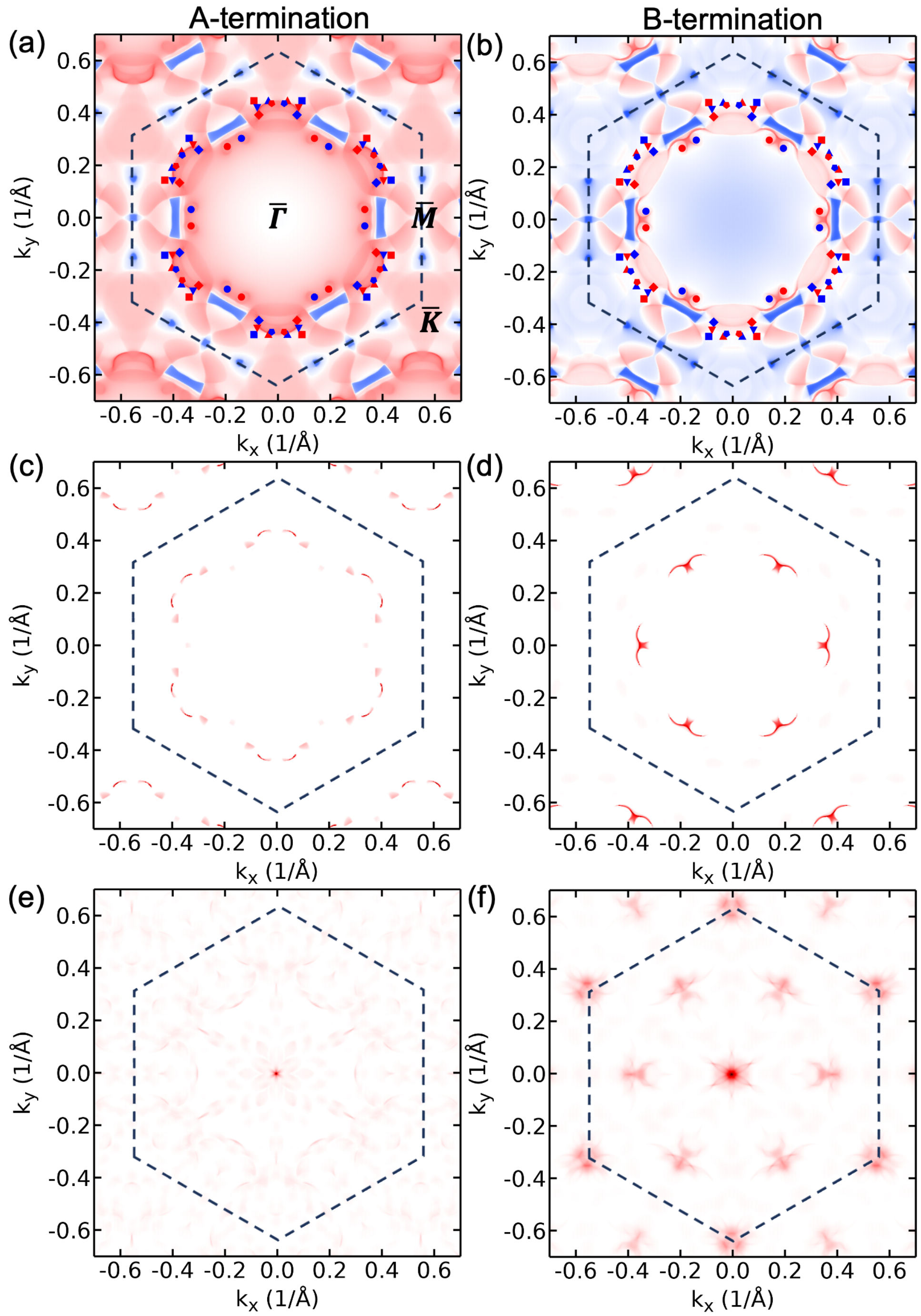

Figure 7. (a-b) 2D Fermi surface of $\gamma$-$PtBi_2$ (0001) at $E_F$+0.100 eV calculated with the averaged experimental structural parameters. The two columns are for the A and B-termination, respectively. (c-d) Fermi surface with only surface state contribution and (e-f) the corresponding QPI. For Fermi surface, red (blue) stands for high (low) intensity. For surface state only contribution and QPI, red (white) stands for high (low) intensity. The multiple sets of Weyl points are labeled in 2D Fermi surface as different point shapes with red (blue) for positive (negative) chirality.